\documentstyle[11pt,newpasp,twoside,epsf]{article}
\def\edcomment#1{\iffalse\marginpar{\raggedright\sl#1\/}\else\relax\fi}
\reversemarginpar

\begin{document}
\title{Symbiotic Miras vs.\ Planetary Nebulae in the Near Infrared}
 \author{Stefan Schmeja \& Stefan Kimeswenger}
\affil{Institut f\"ur Astrophysik, Leopold-Franzens-Universit\"at Innsbruck, 
Technikerstr.\ 25, A-6020 Innsbruck, Austria}

\begin{abstract}
While symbiotic Miras and planetary nebulae are hard to distinguish by optical spectroscopy, their near infrared colors differ. We propose the near infrared two-color diagram to be an excellent tool to easily distinguish these two classes of objects.
\end{abstract}

\section{Introduction}

Nebulae around symbiotic Miras are very similar to planetary nebulae (PNe) in terms of morphology, excitation conditions, and chemical abundances, although they are formed in a slightly different way (Corradi 2002). While they are hard to distinguish by means of optical spectroscopy, their near infrared (NIR) colors differ noticeably.
We obtained NIR photometry of a sample of 155 PNe and (known or suspected) nebulae around symbiotic Miras observed with DENIS (Deep Near Infrared Southern Sky Survey; Epchtein et al.\ 1997) in Gunn-$I$ (0.82~\micron), $J$ (1.25~\micron), and $K{\rm _s}$ (2.15~\micron). The catalog is presented in Schmeja \& Kimeswenger (2002a, 2003), details on the selected sample and on the data reduction are given in Schmeja \& Kimeswenger (2001).

\section{Results and Discussion}
The dereddened NIR two-color diagram of our sample of PNe and symbiotic Miras is shown in Fig.~1. A vector corresponding to a reddening of $E_{\rm B-V}=1$ is also shown, indicating the direction the objects would move in the diagram when  the extinction values are increased.
The symbiotic Miras are clearly separated from the genuine PNe as well as from the classical Miras. So the colours do not only represent the cool component.

\begin{figure}[t]
\centerline{\epsfysize=10cm
\epsfbox{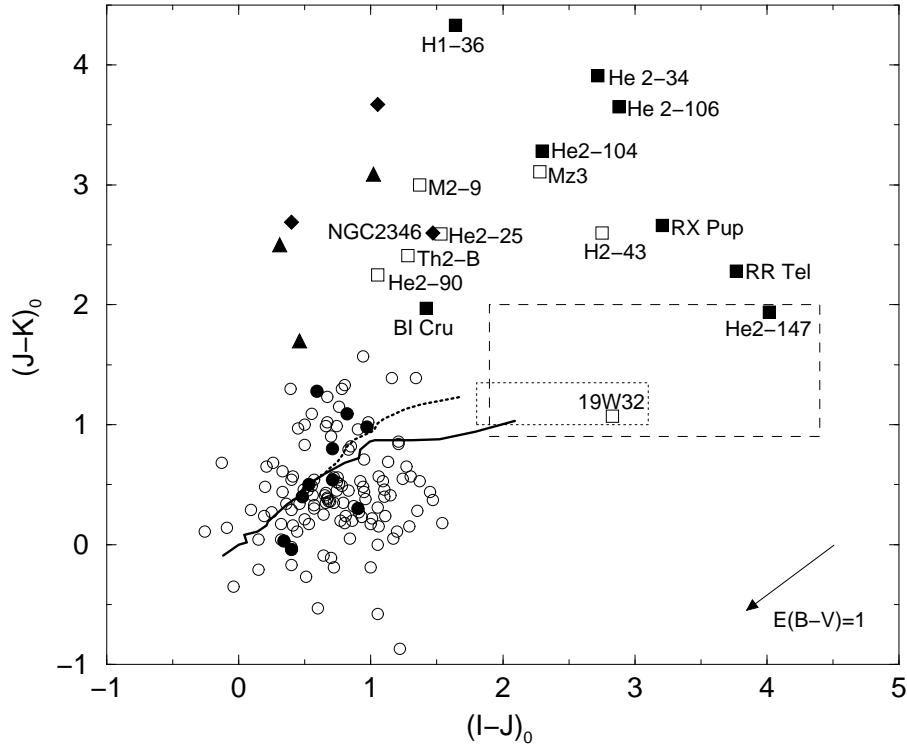}}
\caption{Dereddened $IJK$ diagram of genuine PNe (circles), symbiotic Miras (squares), and suspected symbiotic Miras (open squares). Bipolar PNe with regular colors are marked as filled circles.
Also shown are the positions of the stellar main sequence (solid line), the giants (dotted line), IR-[WC] PNe (triangles), peculiar PNe (diamonds) and classical Miras and semiregular variables (dashed and dotted box, respectively; after Hron \& Kerschbaum (1994) and Whitelock, Marang \& Feast (2000)).}
\label{2colour-diag}
\end{figure}

Whitelock (1987) used a model of a 2500~K star and a 800~K envelope with a circumstellar extinction $0\fm3 \la A_{\rm K} \la 1\fm5$ to explain the $JHK$ colors of symbiotic Miras.
 This simple extinction+cold emission model describes the 1-5\,$\mu$m colors well. Such a circumstellar extinction would lead to a circumstellar reddening of at least the same amount in case of $(I-J)$. However, we measure colors that are even bluer than those of classical Miras, so a modification of this view is needed.
Also the spread in $(I-J)$ is significantly higher than in the $JHK$ two-color diagram.
$(I-J)$ is therefore indicating the strength of the emission originating from ionized gas and the hot component. Those parts cannot be obscured by the circumstellar shell as proposed by the model of Whitelock. 
More sophisticated models are needed to describe the spectral energy distribution (SED). 

All the PNe suspected to be symbiotic Miras (except 19W32) also lie in the mentioned region: Mz~3, M~2-9, H~2-43, He~2-25, Th~2-B, and 19W32 are discussed in detail in Schmeja \& Kimeswenger (2001), He~2-90 has been recently suspected to be a symbiotic by Guerrero et al.\ (2001). Two objects shall be discussed briefly: SaSt~1-1 (AS~201) is a nebula around a symbiotic star, but shows colors like a genuine PN. However, this object is not a symbiotic Mira, but a yellow symbiotic where the cool component is a G5 star. The nebula is most likely the current PN of the white dwarf. NGC~2346 is a bipolar PN placed among the symbiotic Miras in our diagram. Although it has a binary central star, there are no hints that it is a symbiotic system. However it might be a triple system (Phillips \& Cuesta 2000), perhaps with a Mira?
Other types of objects shown in the diagram, like Infrared-[WC] PNe and peculiar PNe are discussed in Schmeja \& Kimeswenger (2002b).

Fig.~2 shows a plot of the dereddened near infrared
magnitudes of the nebulae versus the logarithm of their 5~GHz (6~cm)
radio flux. The radio fluxes are taken from the literature and are listed in our catalog (Schmeja \& Kimeswenger 2002a, 2003).  The lines indicate the predicted
radio-infrared relation for a plasma with a temperature of
10$^4$~K (Whitelock 1985). Most of the genuine PNe lie close to the
theoretical line, in {\it J} there is a small systematic infrared
excess. 
This corresponds well with the results of Whitelock
(1985) and Pe{\~n}a \& Torres-Peimbert (1987). 
Whitelock (1985)
attributes this excess to strong emission in the helium triplet
line at 1.083~\micron. 
As expected, the objects with peculiar
colors in the two-color diagram show an infrared excess,
especially in {\it K}: The symbiotic Miras 
(marked as open squares) show a much higher flux in {\it K} than
expected from their radio flux.
These results underline the conclusions from the two-color diagram that a significantly higher {\it K} flux causes the peculiar colors of those objects.

\begin{figure}[t]
\centerline{\epsfysize=8.5cm
\epsfbox{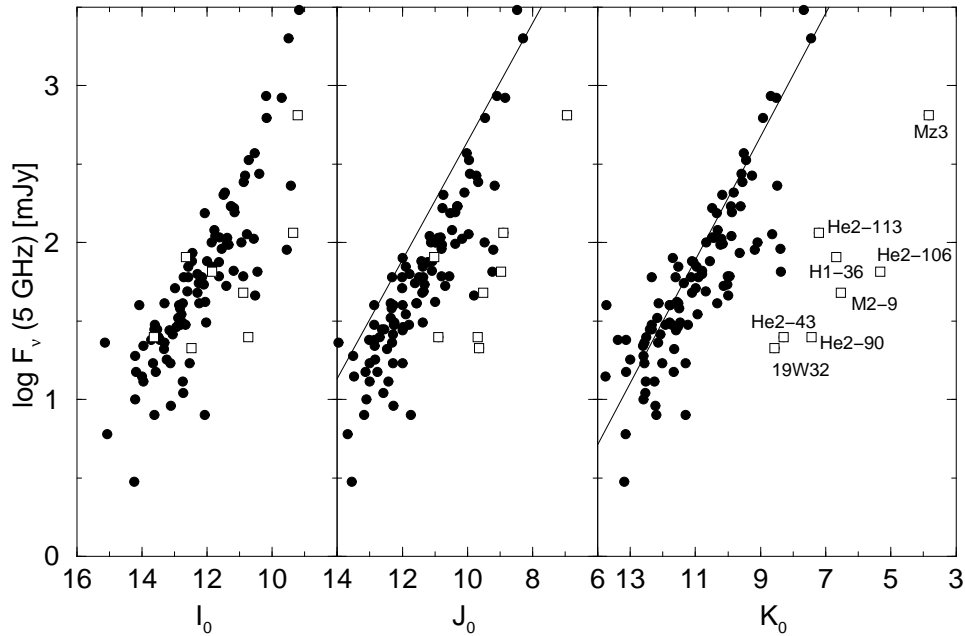}}
\caption{Dereddened NIR magnitudes versus 5~GHz fluxes of genuine PNe (circles) and symbiotic Miras (squares).}
\label{radio}
\end{figure}
\section{Conclusions}
The DENIS NIR two-color diagram is an excellent and easily applicable tool to distinguish nebulae around symbiotic Miras from genuine PNe. Unfortunately, it does not apply to yellow symbiotics, however. Several (but certainly not all) bipolar PNe, like Mz~3 or M~2-9 turn out to be in fact symbiotic Miras. A widely ignored fact is the inconsistent classification of the discussed objects: Whereas many nebulae around symbiotic Miras (e.\,g. He~2-104, He~2-147) are classified as PNe, others, like BI~Cru, that shows a similar nebula, are not. Whether those objects are accepted or rejected as PNe, or whether a new designation like ``symbiotic proto-PNe'' (Kohoutek 2001) is introduced, at least this should be done consistently for all objects of that kind.


\begin{references}
\reference Corradi, R.~L.~M. 2002, ASP Conf.\ Ser.\ (this volume)
\reference Epchtein, N., et al. 1997, The Messenger, 87, 27
\reference Guerrero, M.~A., Miranda, L.~F., Chu, Y.-H., Rodr\'{\i}guez, M., \& Williams, R.~M. 2001, \apj, 563, 883
\reference Hron, J., \& Kerschbaum, F. 1994, \apss, 217, 137
\reference Kohoutek, L. 2001, Catalog of Galactic Planetary Nebulae (updated version 2000), Abhandlungen aus der Hamburger Sternwarte Band XII
\reference Pe\~{n}a, M., \& Torres-Peimbert, S. 1987, Rev.\ Mex.\ Astron.\ Astrofis., 14, 534 
\reference Phillips, J.~P., \& Cuesta, L. 2000, \aj, 119, 335 
\reference Schmeja, S., \& Kimeswenger, S. 2001, \aap, 377, L18
\reference Schmeja, S., \& Kimeswenger, S. 2002a, Ionized Gaseous Nebulae,  Rev.\ Mex.\ Astron.\ Astrofis.\ (Ser.\ de Conf.), 12, 176 
\reference Schmeja, S., \& Kimeswenger, S. 2002b, IAU Symp.\ 209, ASP Conf.\ Ser., in press
\reference Schmeja, S., \& Kimeswenger, S. 2003, Hvar Obs.\ Bull., in press
\reference Whitelock, P.~A. 1985, \mnras, 213, 59
\reference Whitelock, P. A. 1987, \pasp, 99, 573
\reference Whitelock, P., Marang, F., \& Feast, M. 2000, \mnras, 319, 728
\end{references}
\end{document}